\documentstyle[12pt]{article}
\title{Bose-Einstein correlations in multiple particle production}
\author{Kacper Zalewski\thanks{Supported in part by the KBN grant
2P03B 086 14}\\
Jagellonian University\\ and\\ Institute of Nuclear Physics, Krak\'ow, Poland}

\begin{document}
\maketitle

\begin{abstract}

Bose-Einstein correlations are studied in the framework of a class of
independent particle production models. This generalizes the studies for a
variety of models proposed previously. It is shown that the Bose-Einstein
correlations lead for this class of models to Einstein's condensation at
sufficiently high density. They also enhance unusual charge distributions and
may explain the centauro and anticentauro events reported by cosmic ray
physicists. For typical models the correlations cause a shrinking of the
momentum distribution of the produced identical particles and an apparent
shrinking of the production region.

\end{abstract}

\section{INTRODUCTION}

This short communication is based on three papers done in collaboration with
Andrzej Bialas \cite{BZ1},\cite{BZ2},\cite{BZ3}. Further details and references
to earlier work can be found in these papers. Here only the main assumptions
and conclusions will be presented. Bose-Einstein correlations are an apparent
attraction in momentum space between identical bosons. This effect is due to
the symmetrization of the state vectors with respect to exchanges of identical
particles, which is required by Bose-Einstein statistics. Let us mention two
reasons why, among the many known correlations, these correlations are
considered particularly interesting. The first reason is related to the fact
that the Bose-Einstein attraction between particles is strong if and only if
these particles are close to each other in momentum space. Thus, the effect
becomes more and more important, when the density of (identical) particles
increases. The RHIC heavy ion collider will begin taking data soon and there
the particle densities will be very high. Consequently, the Bose-Einstein
correlations are likely to be very significant. Since, as we shall see in the
following, they give a number of effect, which are not intuitively expected,
their understanding will be important for a correct interpretation of the data.
Another reason is that from the Bose-Einstein corrections to the correlation
functions of identical particles it is possible to get information about the
size and shape of the region, where the particles are produced. Similarly,
Hanbury-Brown and Twiss have successfully used the Bose-Einstein correlations
among photons coming from stars to find the radii of these stars. Unfortunately
the simplifying assumptions, which are well justified in the case of
photons emitted by a star, are not always plausible for particles produced in a
high energy scattering process. Thus the results for particles are rather model
dependent. Since, however, no better method of measuring the particle
production region is known, and since information about its size is necessary,
e.g. to answer the question, whether the energy density in this region is big
enough to expect that the quark-gluon plasma is formed there, the study of
Bose-Einstein correlations attracts much interest as a source of information
about the particle production region.

\section{ ASSUMPTIONS AND A SIMPLE MODEL}

We make two assumptions. One is physical and the other is technical, chosen to
make the problem easy to solve. The physical assumption is that intuition is
applicable to the production of distinguishable particles, or perhaps more
realistically to the production of particles at low density, where the
Bose-Einstein correlations are not important. According to this assumption we
guess for distinguishable particles a density matrix $\rho_0(q,q')$. Here
$q(q')$ is the set of all the momentum components of all the particles being
produced. Thus for the production of $N$ particles vector $q$ has $3N$
components. The diagonal elements of the density matrix give, as usual, the
momentum distribution

\begin{equation}
\Omega_0(q) = \rho_0(q,q).
\end{equation}
It is assumed that the trace of the density matrix $\rho_0$, or equivalently
the integral over the momentum distribution over all the $3N$-dimensional
momentum space, is one. In order to make numerical predictions it is
necessary to assume some functional form for $\rho_0(q,q')$, but here we will
be interested mostly in general results, thus no specific function is
introduced.  For the production of $N$ identical bosons formula (1) should be
replaced by the symmetrized formula

\begin{equation}
\Omega_N(q) = \sum_P \mbox{Re } \rho_{0N}(q,q_p),
\end{equation}
where $q_P$ is the $3N$ dimensional momentum vector, which arises from the
vector $q$, when permutation $P$ is applied to the ordering of the
$N$ particles. The summation extends over all the $N!$ permutations of the $N$
identical bosons.

Formula (2) has important implications. As our first example let us integrate
both sides of equality (2) over $q$. Thus we obtain the integrated
cross-section for the $N$ identical bosons:

\begin{equation}
\sigma_N = \int dq\;\Omega_N(q) = 1 + \ldots.
\end{equation}
According to formula (2) the integral is a sum of $N!$ terms. The first term
corresponding to the identical permutation $P=1$ gives one because of our
normalization condition. The other terms, however, change $\sigma_N$. Since the
change depends on the particle number $N$, the multiplicity distribution of the
produced particles is also changed. In order to go further, we must introduce
some technical assumption to make the calculation feasible. According to our
general philosophy, we make the assumptions for distinguishable particles and
then we go over to the indistinguishable particles by symmetrizing. We assume
that the particles are produced independently in the sense that the
multiplicity distribution is poissonian

\begin{equation}
P_0(N) = \frac{\nu^N}{N!} e^{-N}.
\end{equation}
For each multiplicity, moreover, we assume that the state of the produced
particles is pure i.e. that $\rho_{0N} = |\psi_N\rangle \langle \psi_N|$. This
is a rather unrealistic assumption and soon we will replace it by something
better, but as we shall see the present simple case illustrates important
features of the more realistic model in a very simple way. Since the state
vector of $N$ particles does not change under permutations of identical
bosons, we find for the integrated cross-section of producing $N$
indistinguishable particles

\begin{equation}
\sigma_N = N! \sigma_{0N}.
\end{equation}
Consequently the corrected (unnormalized) probability for producing $N$
particles is by a factor $N!$ larger than for the distinguishable particles
and, if the probability distribution for the distinguishable particles is given
by the poissonian distribution (4), for the indistinguishable particles we have
the (normalized) probability distribution

\begin{equation}
P(N) = (1-\nu) \nu^N,
\end{equation}
which is a geometrical distribution, very different from the original
poissonian one. Let us note two implications of this formula. When many pions
are produced, usually about one third of them are $\pi^0$-s. If the number of
$\pi^0$-s is governed by the poissonian distribution, large deviations from
this average are very unlikely. For the geometric distribution, however, large
deviations from the average are much more probable. This could explain the
centauro and anticentauro events reported by cosmic ray physicists. In such
events the $\pi^0$-s are respectively either completely absent, or they are the
only pions produced (in the region of momentum space accessible to experiment!).
Another observation is that according to distribution (6) the average
multiplicity of particles is

\begin{equation}
\overline{N} = \frac{\nu}{1-\nu}.
\end{equation}
For small values of $\nu$ this reproduces the result of the poissonian
distribution as expected. At $\nu = 1$, however, there is a singularity. In the
more realistic model introduced below this corresponds to the Einstein
condensation.

\section{A BETTER MODEL}

A more realistic model can be obtained by replacing the assumption of a pure
state by the assumption that at given $N$ the density matrix for
distinguishable particles is a product of single particle density matrices. Thus

\begin{equation}
\rho_{0N} = \prod_{i=1}^N \rho_{01}(q_i,q_i'),
\end{equation}
where $q_i (q_i')$ is the momentum vector of particle $i$. Also for this model
all the calculations can be easily performed. This model is not yet realistic,
but it is certainly much closer to reality than the previous one and some of
its implications may be relevant for real experiments. We will present four
predictions.

The single particle momentum distribution, the two-particle correlation
function and all the $n$-particle correlation functions for $n=3,4,\ldots$ can
be expressed in terms of one function $L(q_1,q'_1)$. Thus (if it is legitimate
to neglect the variations of phase of the function $L$) one can measure the
two-body correlation function and from that predict without further assumptions
all the other correlation functions. Our model is too crude to expect  more
than qualitative agreement with experiment, but the formulae exist and can be
checked.

The following two predictions do not hold for an arbitrary choice of the
density matrix $\rho_{01}$, but it is plausible that they hold for "reasonable"
choices. Firstly, the momentum distribution is expected to shrink as compared
to the unsymmetrized distribution. Suppose that centauro, or anticentauro,
events are found in accelerator experiments. This would mean that the
Bose-Einstein correlations for these events are very large. We predict that the
relevant groups of charged (or neutral) pions should have reduced relative
momenta as compared e.g. to Monte Carlo calculations, which do not include
symmetrization. Secondly, the radius of the production region, as calculated
using standard methods, decreases. This can serve to illustrate our point of
view that intuition should be applied to the uncorrelated case. Suppose that
the pions are produced in a sphere the size of a nucleus. When the density of
the produced particles in momentum space increases, the measured production
radius decreases, although "really" the particles are produced all the time
from the same spherical region. For this reason we propose to interpret the
radius measured at low particle density as the "true" radius, while the radii
measured at higher densities are only "effective" radii.

Finally, we expect the Einstein condensation at sufficiently high density. Out
of the many ways of exhibiting this phenomenon let us look at the formula for
the function $L(q,q')$, which can be rewritten in the form

\begin{equation}
L(q,q') = \frac{\psi_0(q)\psi_0^*(q')}{1 - \nu \lambda_0} + \tilde{L}(q,q').
\end{equation}
Here $\psi_0$ is the eigenfunction of the single particle density operator,
which corresponds to the largest eigenvalue $\lambda_0$. The parameter $\nu$ is
that from the poissonian distribution $P_0(N)$. When $\nu \lambda_0 \rightarrow
1$, the term $\tilde{L}$ stays bounded, while the first term tends to infinity.
At sufficiently high density, $\tilde{L}$ becomes negligible and almost all the
particles are in the state $|\psi_0 \rangle$. This is Einstein's condensation
and also the case considered in the framework of our first simplified model.


\begin{thebibliography}{9}
\bibitem{BZ1}A. Bialas and K. Zalewski, hep-ph/9803408 and Eur. J. Phys. C in
print.
\bibitem{BZ2}A. Bialas and K. Zalewski, hep-ph/9806435 and Phys. Letters B in
print.
\bibitem{BZ3}A. Bialas and K. Zalewski, hep-ph/9807382 and Slovak J. Phys in
print.

\end{thebibliography}
\end{document}